# Gap Cycling for SWIFT

Curtis A. Corum*, Djaudat Idiyatullin, Carl J. Snyder, and Michael Garwood

Center for Magnetic Resonance Research
Dept. of Radiology, Medical School,
University of Minnesota
Minneapolis, Minnesota, USA

*Correspondence:
Curtis A. Corum, Ph.D.
Assistant Professor of Radiology
University of Minnesota, Medical School
Center for Magnetic Resonance Research
2021 6th St. SE
Minneapolis, MN 55455

corum@cmrr.umn.edu
612-625-8258 office
612-626-2004 fax

**Abstract**

**Purpose:** SWIFT (SWeep Imaging with Fourier Transformation) is a non-Cartesian MRI method with unique features and capabilities. In SWIFT, radiofrequency (RF) excitation and reception are performed nearly simultaneously, by rapidly switching between transmit and receive during a frequency-swept RF pulse.

Because both the transmitted pulse and data acquisition are simultaneously amplitude-modulated in SWIFT (in contrast to continuous RF excitation and uninterrupted data acquisition in more familiar MRI sequences), crosstalk between different frequency bands occurs in the data. This crosstalk leads to a "bulls-eye" artifact in SWIFT images.

We present a method to cancel this inter-band crosstalk by cycling the pulse and receive gap positions relative to the un-gapped pulse shape. We call this strategy "gap cycling."

**Methods:** We carry out theoretical analysis, simulation and experiments to characterize the signal chain, resulting artifacts, and their elimination for SWIFT.

**Results:** Theoretical analysis reveals the mechanism for gap-cycling's effectiveness in canceling inter-band crosstalk. We show phantom and *in-vivo* results demonstrating bulls-eye artifact free images.

**Conclusion:** Gap cycling is an effective method to remove bulls-eye artifact resulting from inter-band crosstalk in SWIFT data.

## Introduction

SWIFT (SWeep Imaging with Fourier Transform) (1) utilizes gapped frequency-swept pulses (2) for excitation; signal is received after a short dead interval. SWIFT can be categorized as one of the general class of short-$T_2$ or $T_2^*$ sensitive sequences (3) which possess a very short time interval between signal excitation and reception. For SWIFT this time is on the order of microseconds (Fig. 1).

Due to the gapping (gating on and off) of both the transmit pulse and the receiver acquisition interval, some signal artifacts are present in SWIFT data. This class of artifactual signal was first noted as baseline or phantom peaks with homonuclear decoupling applied during gapped signal reception. In SWIFT this artifactual signal manifests as a spherically symmetric "bulls-eye" artifact (Fig. 2A) and additional noisy background intensity in images (4,5).

The amount of bulls-eye artifact is object, as well as SWIFT parameter dependent. As noted in reference (2), the bulls-eye artifact in SWIFT can be caused by errors in the pulse profile, which when corrected, result in less artifact. We have previously described correction algorithms for reducing bulls-eye artifact, irrespective of the source (4,5).

In the following we discuss the dominant remaining component of the bulls-eye artifact, inter-band crosstalk and introduce a method called "gap cycling" for cancelling the artifact in SWIFT images.

## Theory

SWIFT's transmit pulse can be viewed as an amplitude-modulated (by a square wave pattern) version of the un-gapped pulse (2). The modulation creates sidebands which extend beyond the base profile bandwidth (labeled Band 0) of the un-gapped pulse (Fig. 3B). These sidebands can excite signal from spatial (or frequency) regions beyond the expected field of view. With SWIFT, the spin signal is the time domain convolution of the pulse with the spin impulse response (eq. 1) which becomes multiplication in the frequency domain. In addition, off-resonance nutation signal is present (6) that extends to the entire bandwidth of the pulse, including the gapping sidebands (Fig. 3D). This extension of the spin signal bandwidth is present (to a lesser extent) even with an un-gapped pulse (Fig. 3C).

Following the notation in (2), with the impulse response of a 1-d object (Fig. 3A shows frequency domain representation) expressed as $h(t)$ and the excitation pulse as $x(t)$, we have for the excited signal $r(t)$:

$$r(t) = h(t) \oplus x(t), \qquad \text{Eq. 1}$$

where $\oplus$ is the convolution operation.

We substitute $x_{gap}(t)$ for $x(t)$, where

$$x_{gap}(t) = x(t)\, s_p(t) \quad \text{and} \qquad \text{Eq. 2a}$$

$$s_p(t) = \text{comb}[t/(\lambda\, \Delta t)] \oplus \text{rect}[(t - \delta t)/(\rho_p\, \lambda\, \Delta t)], \qquad \text{Eq. 2b}$$

with the duty cycle of the pulse given by $\rho_p$, and the constants $\lambda\, \Delta t$ represent the period between gaps in the synthesized pulse, where $\lambda$ is the integer oversampling parameter and $\Delta t$ is the synthesis interval.

We have introduced:

$$\delta t = m'\, \lambda\, \Delta t \ \ \text{with}\ \ m' \equiv m/M,\ \ 0 \le m \le M-1, \ \ \text{and}\ \ 0 \le M \le \lambda \qquad \text{Eq. 2c}$$

which represents a shift in the position of the pulse gaps. $M$ is a convenient integer value which represents the number of possible positions of the gap, and $m$ is an integer representing the position. An additional constraint on $M$ is that it should be commensurate with $\lambda$.

For simplicity we will use the same parameters for the receiver except to allow a different duty cycle $\rho_r$ and neglect the delay between the center of the pulse and receiver intervals for which we compensate by linear phase correction of the data.

The gated receiver in SWIFT can also be thought of as an amplitude-modulated version of a continuous receiver. We capture the gapped receiver's effect on the data $d(t)$ by the equation:

$$d(t) = r(t)\, s_r(t), \qquad \text{Eq. 3}$$

with $s_r(t) = \text{comb}[(t)/(\lambda\, \Delta t)] \oplus \text{rect}[(t - \delta t)/(\rho_r\, \lambda\, \Delta t)]$.

Substituting Eq. 1 into Eq. 3, we have:

$$d(t) = \{\, h(t) \oplus [x(t)\, s_p(t)]\}\, s_r(t), \qquad \text{Eq. 4}$$

which captures the effects on the data of both the gapped pulse and gapped receive interval.

It is more convenient to analyze Eq. 4 in the frequency domain. After Fourier transforming and using the Fourier convolution theorem we obtain:

$$D(f) = \{H(f)[X(f) \oplus S_p(f)]\} \oplus S_r(f), \qquad \text{Eq. 5}$$

with $S_p(f) = \rho_p\, \lambda^2 \Delta t^2\, \text{comb}(\lambda\, \Delta t\, f)\, \text{sinc}(\rho_p\, \lambda\, \Delta t\, f)\, e^{i2\pi\, m'\lambda\, \Delta t\, f}$,

and $S_r(f) = \rho_r\, \lambda^2 \Delta t^2\, \text{comb}(\lambda\, \Delta t\, f)\, \text{sinc}(\rho_r\, \lambda\, \Delta t\, f)\, e^{i2\pi\, m'\lambda\, \Delta t\, f}$.

The data spectrum $D(f)$ is shown (Fig. 3E) and consists of the spin spectrum (Fig. 3D) with overlapped multiple replicas due to the receiver modulation. Subtraction of the scaled spin spectrum from $D(f)$ shows the overlapping components remaining in the baseband data spectrum (Fig. 3F.) This remaining overlapping signal (Fig. 3F, Band 0) is the origin of the bulls-eye artifact.

Analysis of Eq. 5 is accomplished by looking at each frequency band of size $sw$, where $sw = 1/(\lambda\, \Delta t)$.

We define crosstalk as signal originating from one band becoming mixed into

Gap Cycling for SWIFT 5

other bands. We can evaluate the effects of $S_p(f)$ and $S_r(f)$ by looking at the components of the data spectrum:

$$D_{j,k}(f - f_k) \equiv H(f - f_j) X(f) C_{j,k}, \qquad \text{Eq. 6}$$

where $C_{j,k}$ captures the effect of convolution by $S_p(f)$ and $S_r(f)$; integer $j$ indexes the signal originating from the corresponding pulse band centered at $f_j = j\, sw$; and $k$ indexes the signal received in the corresponding band of the received data at $f_k = k\, sw$. For example, $j = k = 0$ corresponds to the signal excited by the central band of the pulse (the "pulse baseband") and received in the central band of the receiver (the "receiver baseband"). We assume for simplicity in this analysis that the un-gapped pulse shape $X(f)$ is band limited: $X(f)$ lies within the bandwidth $sw$, as currently implemented in SWIFT.

The components $C_{j,k}$ define a "crosstalk matrix." When non-zero off-diagonal elements exist, there is mixing between bands, resulting in artifact. We evaluate Eq. 5 and put into the form of Eq. 6 and obtain:

$$C_{j,k} \equiv \lambda^4 \Delta t^4 \rho_p \rho_r \operatorname{sinc}(\rho_p\, j) \operatorname{sinc}[\rho_r\,(k - j)]\, e^{i2\pi\, m'(k-j)}. \qquad \text{Eq. 7}$$

We note immediately that when $\rho_r = 1$ the second sinc term is only non-zero at the center, corresponding to $j = k$, in which case $C_{j,k}$ forms the components of a diagonal matrix. This is the no-crosstalk situation familiar from conventional continuous acquisition MRI. In the situation with gapped SWIFT, in which $\rho_p < 1$ and $\rho_r < 1$, we have non zero off-diagonal terms.

**Methods**

We have developed a scheme (gap cycling) where the position of gaps in the pulse and receiver are cycled relative to the un-gapped pulse shape during repeated TR periods. The pulse shape is shifted, before each TR period, relative to the position of the transmitter and receiver gates, with the intention of providing data to cancel crosstalk. We note that for the receiver baseband, $k = 0$, we can write:

$$\overline{D_{J,0}}(f) = H(f - f_j) X(f) \sum_{m=0}^{M-1} C_{j,0,m} \qquad \text{Eq. 8}$$

which represents the cycle through the integer values of $m$ from $0 \leq m \leq M - 1$ and averaging the result. Note we have added the dependence on $m$ to to the crosstalk matrix component, through $m'$. Evaluating, we have:

$$\sum_{m=0}^{M-1} C_{j,0,m} = \lambda^4 \Delta f^4 \rho_p \rho_r \, \text{sinc}(\rho_p \, j) \sum_{m=0}^{M-1} e^{-i2\pi \, m' j} \qquad \text{Eq. 9}$$

in which the sum on the right is zero unless $j = 0$, since the exponential term moves evenly through a full cycle of the complex unit circle. Hence, the contribution from all other bands $j \neq 0$ has been cancelled.

We acquired high resolution SWIFT images with two variations of gap cycling and with no gap cycling to evaluate effectiveness. The object was a breast phantom (7) placed in our single-breast coil (8), which was modified to be SWIFT-compatible by removal of the thermoplastic basket. All images were collected with 62.5 kHz baseband bandwidth ($sw = 62.5 \, kHz$), and utilize a Halton-sorted Saff vieworder (9,10) with 65,536 radial spoke views uniformly covering the 3-d sphere of k-space. The excitation pulse is HS2 R64 (2), yielding on average 32 k-space points followed by 192 additional un-gapped (full receive duty cycle) acquisition points. The images are reconstructed by gridding (11,12) from 224 post correlation (1) k-space radial points to a $512^3$ image matrix. One modification to the correlation procedure in (1) is that the gap cycled data at cycle position $m$ is correlated to the shifted pulse shape corresponding to $m$. This produces projections with consistent linear phase in the baseband. The images have a FOV of 256 mm giving the reconstructions a nominal 0.5 mm isotropic resolution. These SWIFT images were obtained using our 4 T Agilent/Siemens/Oxford research scanner using VnmrJ 3.2 and CMRRpack SWIFT software (http://www.cmrr.umn.edu/swift). Interleaved CHESS fat suppression was applied every 16 views (13).

In addition, brain images were acquired from a healthy volunteer using the same SWIFT parameters described above. We utilized a multi-channel TEM/Stripline (14,15) transceiver array specifically designed to be SWIFT compatible (16). The individual coil elements of the array were 200 mm in length and 50 mm in width.

A 12.7 mm copper center conductor was separated from the 50 mm ground conductor by a 12.7 mm polytetraflouroethylene (PTFE) dielectric. To eliminate coil background signal from short $T_2$ materials visible with SWIFT (16), a PTFE enclosure was designed to house the individual coil elements. The coil dielectrics were machined such that the rolled copper foil conductors could be press fitted, eliminating the use of adhesive.

Each coil was individually tuned to the proton Larmor frequency at 4T (169.26 MHz) and matched to a 50-ohm coaxial cable. A single 8 kW power amplifier was used in conjunction with an eight-way equal-amplitude splitter for transmit. A geometric phase distribution was used to ensure a circularly polarized transmit field. Images were reconstructed using root sum of squares combination of the data from each of the 8 receive channels.

**Results**

The addition of gap cycling to SWIFT experiments has largely eliminated bulls-eye artifact and the need for post-processing to remove residual artifact. The uncorrected image appears in Figure 4A. Figure 4B provides the result of acquiring each k-space view 16 times with cycle $M = 16$, taking 16 times as long as the image in Fig. 4A. It does have higher SNR than Fig 4A from the effects of averaging. The bulls-eye artifact is completely absent and no dc correction has been applied. Figure 4C utilizes rapid gap cycling with the same acquisition time as Figure 4A. In rapid gap cycling, the cycle is applied to each view in succession without repeating any view directions. The oversampling of k-space near the origin is enough to average out the crosstalk at the spatial frequencies present. Figure 4D is post processing corrected from the same data as 4A, for comparison.

Figure 5 is an *in vivo* brain image taken with gap cycling and no other bulls-eye correction. No bulls-eye artifact is detectable in the image.

**Discussion**

Gap cycling eliminates bulls-eye artifact resulting from the doubly gapped (gapped pulse and gapped receiver) SWIFT signal chain (17). As mentioned previously, other errors can also cause a spherically symmetric bulls-eye artifact, including coil ringdown (18), short $T_2$ signal from the coil (16), pulse fidelity errors (2), and gradient timing (19), and will not necessarily be addressed by the gap cycling method. We have reason to believe that residual pulse fidelity errors not addressed by the methods in (2) can be corrected by gap cycling, especially if the number of cycle positions *M* is comparable to the pulse synthesis oversampling parameter λ, but a firm conclusion requires further work.

As noted previously, gap cycling is not the only way to remove or reduce bulls-eye artifact in SWIFT data. The post-processing method (4) can significantly reduce bulls-eye artifact, but can also introduce similar artifact depending on object structure. We have extended the procedure to minimize object-dependent artifact (5), but this procedure still leaves residual artifact coinciding with radii where the object has significant edge transition (Fig. 2B).

Since gap cycling does not require post processing correction, it avoids potential noise amplification and bias (due to regularization) inherent in inversion-based -processing. Additional artifact introduced by the post-processing method can be viewed as due to a bias term introduced by the regularization process and/or modeling error (20). Inversion schemes are also utilized in the methods proposed in the references (21,22).

To date, rapid gap cycling, where the gap cycle is applied to successive views, has demonstrated nearly the same performance in reducing bulls-eye artifact as the full cycle applied to each view. We theorize this is due to the spatial frequencies present in the artifact being significantly lower than the resolution of the image. The oversampling at the center of k-space even in an under-sampled 3-d radial acquisition is sufficient to provide each post gridding k-space point with a full cycle. Further work is needed to establish the conditions under which rapid gap cycling breaks down, but empirically it has worked for a wide range of view orders, objects, degree of under sampling, and SWIFT pulse sequence parameters.

**Conclusion**

We have presented the gap-cycling method to cycle the position of the RF transmit and receive intervals relative to the pulse shape for SWIFT. Gap cycling cancels crosstalk between frequency bands which results from simultaneous amplitude modulation of both the transmitted pulse and receiver interval. Gap cycling completely removes spherically symmetric bulls-eye artifact due to inter-band crosstalk, greatly improving the quality of SWIFT images.

**Acknowledgements**

We gratefully acknowledge NIH grant support from UL1 TR000114, R21 CA139688, S10 RR023730, S10 RR027290, P41 RR008079 and P41 EB015894. We thank Drs. Steve Suddarth and Alan Rath of Agilent for technical support. We thank Mr. Michael Tesch and Dr. Ryan Chamberlain for their collaboration on developing low level SWIFT psg code for the Agilent console. We also thank Ms. Katrina Edenfeld for detailed proofreading and editing, and Dr. Silvia Mangia for suggestions on the manuscript.

**Figure 1**

A. Schematic SWIFT sequence consisting of gapped frequency swept pulse.

B. Enlargement of one period of the gap pattern showing the transmit interval, dead time interval, and receiver interval.

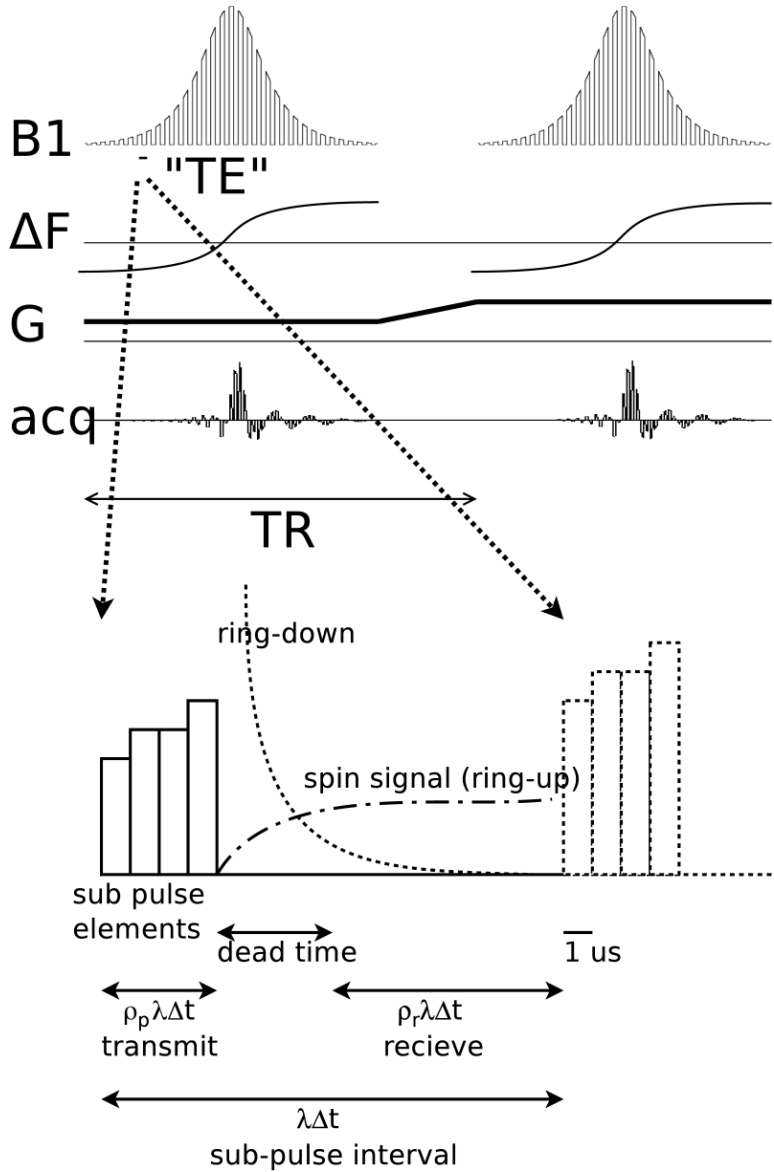

**Figure 2**

A. Uncorrected breast phantom image with intensity scaled to clearly show bulls-eye artifact of concentric rings in the slice (which are spherical shells in 3-d).

B. Corrected breast phantom image with algorithm in references (4,5). The artifact is greatly reduced but still present.

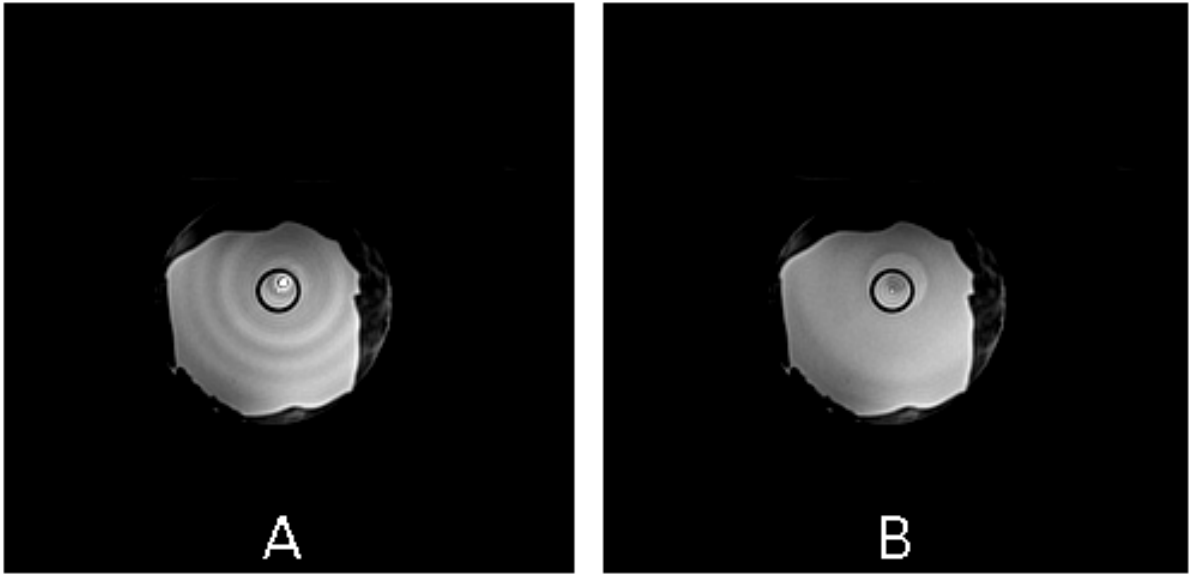

**Figure 3**

A. Impulse Spectrum $H(f)$  
B. Gaped Pulse Profile $X_{gap}(f)$

C. Un-gapped pulse Spin Spectrum $H(f)\,X(f)$  
D. Spin Spectrum $R(f) = H(f)\,X_{gap}(f)$

E. Data Spectrum $D(f)$  
F. Data Spectrum $D(F)$ with scaled $R(f)$ subtracted

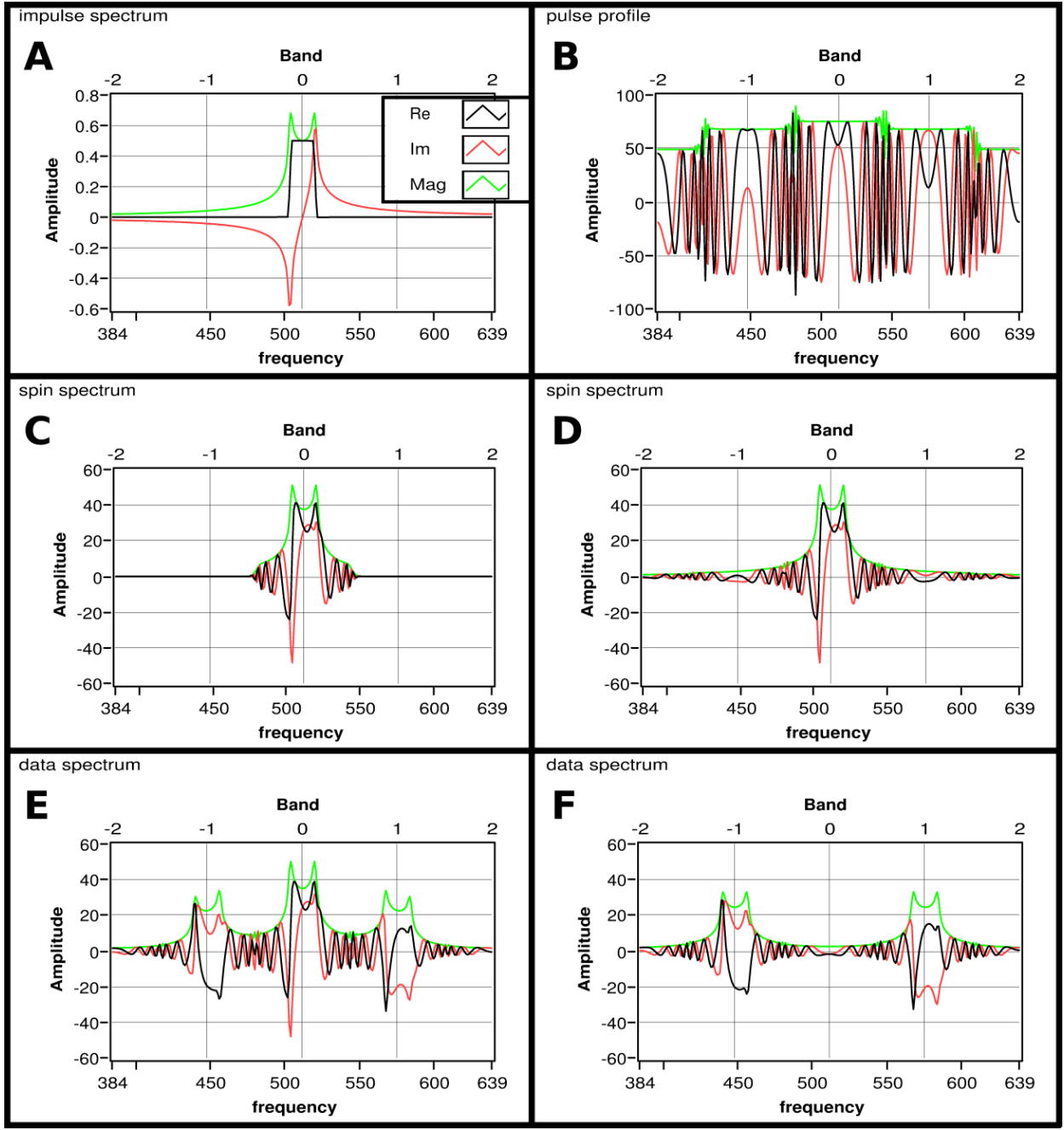

**Figure 4**

A. High resolution SWIFT image, $512^3$ matrix. No gap cycling, no correction.

B. High resolution SWIFT image, $512^3$ matrix. Full gap cycling using 16 acquisitions for each projection (k-space line). Bulls-eye artifact is completely eliminated.

C. High resolution SWIFT image, $512^3$ matrix. Rapid gap cycling, which takes the same time as image A. Each projection is taken only once, but gap cycling is updated in sequence. Bulls-eye artifact is completely eliminated.

D. Algorithm (4,5) applied to dataset from Figure 4A, object dependent artifact remains.

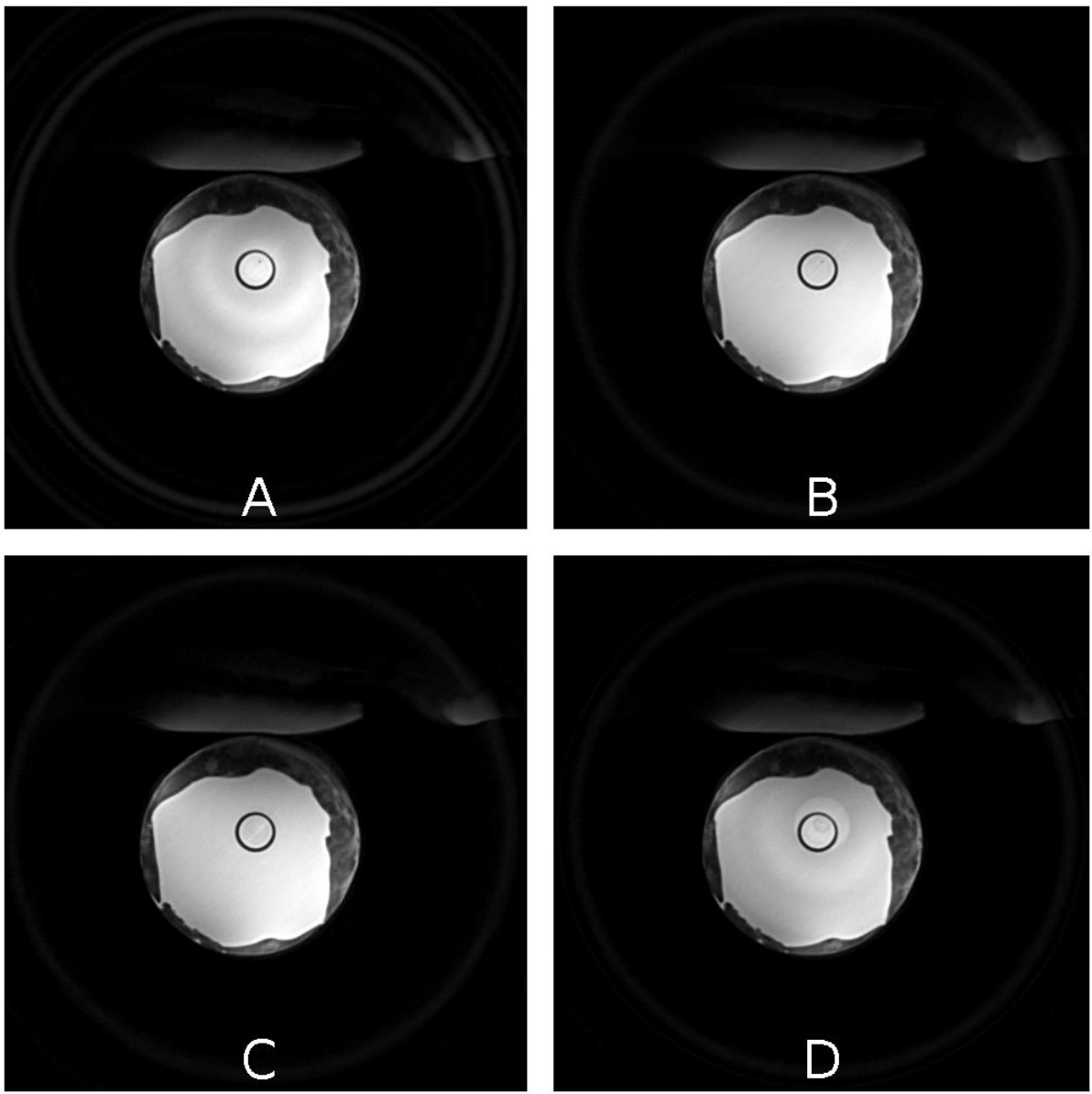

**Figure 5**

High-resolution brain image from SWIFT scan, 320 mm FOV, and $512^3$ matrix yielding 0.625 mm isotropic resolution utilizing rapid gap cycling. No bulls-eye artifact is detectable.

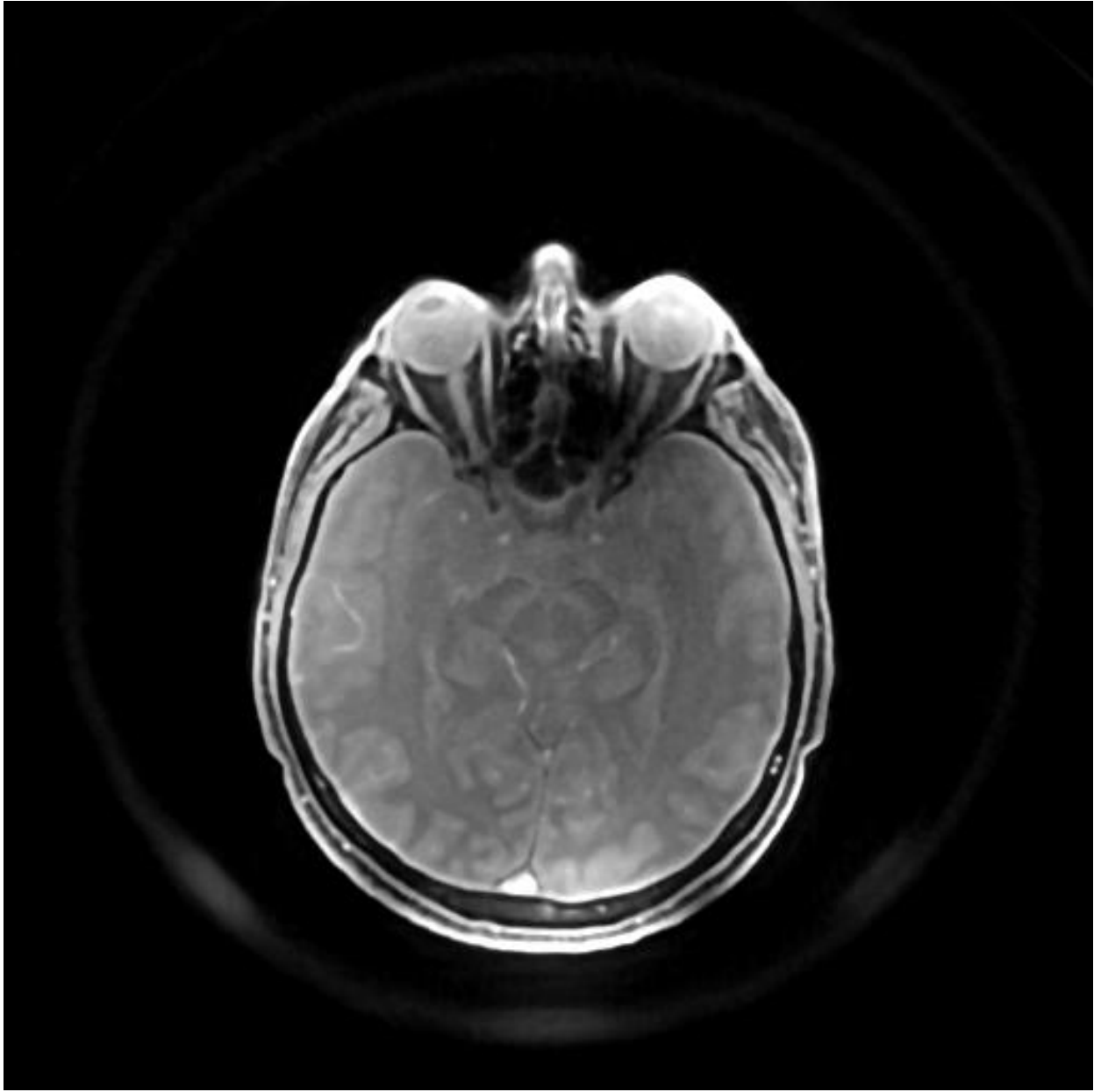

# References

1. **Idiyatullin D, Corum C, Park JY, Garwood M. Fast and quiet MRI using a swept radiofrequency. J Magn Reson 2006;181:342-349.**
2. **Idiyatullin D, Corum C, Moeller S, Garwood M. Gapped pulses for frequency-swept MRI. J Magn Reson 2008;193(2):267-273.**
3. **Wehrli FW. Magnetic resonance of calcified tissues. J Magn Reson 2013;229:35-48.**
4. **Moeller S, Corum C, Idiyatullin D, Chamberlein R, Garwood M. Correction of RF pulse distortions, with application in radial imaging using SWIFT. 2008; Toronto. International Society for Magnetic Resonance in Medicine. p 229.**
5. **Corum CA, Idiyatullin DS, Moeller S, Garwood MG; MULTISTEP CORRECTION FOR ANGLE CONSISTENT ARTIFACTS IN RADIAL SAMPLED DATASETS patent US 20130084023 A1. 2013.**
6. **Levitt MH. Spin Dynamics: basics of nuclear magnetic resonance: Wiley; 2001.**
7. **Bolan PJ, DelaBarre L, Baker EH, Merkle H, Everson LI, Yee D, Garwood M. Eliminating spurious lipid sidebands in 1H MRS of breast lesions. Magnetic Resonance in Medicine 2002;48(2):215-222.**
8. **Merkle H, DelaBarre L, Bolan PJ, Baker EH, Everson LI, Yee D, Garwood M. Transceive Quadrature Breast Coils and Applications at 4 Tesla. ISMRM. Volume 9; 2001. p 1114.**
9. **Saff EB, Kuijlaars ABJ. Distributing many points on a sphere. Mathematical Intelligencer 1997;19(1):5-11.**
10. **Chan RW, Plewes DB. Halton Sequences for Adaptive 3D Radial Sampling. ISMRM Workshop on Data Sampling and Image Reconstruction. Sedona, AZ; 2009.**
11. **Jackson JI, Meyer CH, Nishimura DG, Macovski A. Selection of a convolution function for Fourier inversion using gridding. IEEE Transactions on Medical Imaging 1991;10(3):473-478.**
12. **Pipe JG, Menon P. Sampling density compensation in MRI: rationale and an iterative numerical solution. Magn Reson Med 1999;41(1):179-186.**
13. **Corum CA, Idiyatullin D, Snyder A, Snyder C, Hutter D, Everson L, Eberly L, Nelson M, Garwood M. SWIFT Dual Breast Imaging Sequence and Coil with Interleaved Adiabatic Fat Suppression. ISMRM 21st Annual Meeting. Volume 21; 2013. p 3369.**
14. **Vaughan JT, Hetherington HP, Otu JO, Pan JW, Pohost GM. High frequency volume coils for clinical NMR imaging and spectroscopy. MAGNETIC RESONANCE IN MEDICINE 1994;32(2):206-218.**
15. **Adriany G, Van de Moortele P-F, Wiesinger F, Moeller S, Strupp JP, Andersen P, Snyder C, Zhang X, Chen W, Pruessmann KP, Boesiger P, Vaughan T, Ugurbil K. Transmit and receive transmission line arrays for 7 Tesla parallel imaging. MAGNETIC RESONANCE IN MEDICINE 2005;53(2):434-445.**
16. **Marjanska M, Waks M, Snyder CJ, Vaughan JT. Multinuclear NMR investigation of probe construction materials at 9.4T. Magn Reson Med 2008;59(4):936-938.**
17. **Corum C, Idayatullin D, Moeller S, Garwood M. Doubly-gated (excitation and acquisition) effects in SWIFT. ISMRM Workshop on Data Sampling and Image Reconstruction. Sedona, AZ; 2009.**
18. **Hoult DI. Fast recovery, high sensitivity NMR probe and preamplifier for low frequencies. Review of Scientific Instruments 1979;50(2).**
19. **Peters DC, Derbyshire JA, McVeigh ER. Centering the projection reconstruction trajectory: Reducing gradient delay errors. Magnetic Resonance in Medicine 2003;50(1):1-6.**
20. **Hansen PC. Discrete Inverse Problems: Insight and Algorithms: Society for Industrial and Applied Mathematics (SIAM, 3600 Market Street, Floor 6, Philadelphia, PA 19104); 2010.**

21. Weiger M, Hennel F, Pruessmann KP. Sweep MRI with algebraic reconstruction. Magn Reson Med 2010;64(6):1685-1695.
22. Corum CA, Moeller S, Idiyatullin DS, Chamberlain R, Garwood MG; SIDEBAND PROCESSING FOR MAGNETIC RESONANCE patent US20130043867. 2013.